# Stacking sequence determines Raman intensities of observed interlayer shear modes in 2D layered materials – A general bond polarizability model

Xin Luo, [⊥,£,&]   Chunxiao Cong,[†]   Xin Lu,[†]   Ting Yu,[†,⊥]   Qihua Xiong[†,Ŧ]   and   Su Ying Quek [⊥,£,&,*]

[⊥]Centre for Advanced 2D Materials and Graphene Research Centre
National University of Singapore, 6 Science Drive 2, Singapore 117546

[£]Department of Physics, National University of Singapore, 2 Science Drive 3, Singapore 117551

[&]Institute of High Performance Computing, 1 Fusionopolis Way, #16-16 Connexis, Singapore 138632

[†]Division of Physics and Applied Physics, School of Physical and Mathematical Sciences, Nanyang Technological University, Singapore 637371

[Ŧ]NOVITAS, Nanoelectronics Centre of Excellence, School of Electrical and Electronic Engineering, Nanyang Technological University, Singapore 639798

*To whom correspondence should be addressed. Email: phyqsy@nus.edu.sg

## ABSTRACT

2D layered materials have recently attracted tremendous interest due to their fascinating properties and potential applications. The interlayer interactions are much weaker than the intralayer bonds, allowing the as-synthesized materials to exhibit different stacking sequences, leading to different physical properties. Here, we show that regardless of the space group of the 2D material, the Raman frequencies of the interlayer shear modes observed under the typical $\bar{z}(xx)z$ configuration blue shift for AB stacked materials, and red shift for ABC stacked materials, as the number of layers increases. Our predictions are made using an intuitive bond polarizability model which shows that stacking sequence plays a key role in determining which interlayer shear modes lead to the largest change in polarizability (Raman intensity); the modes with the largest Raman intensity determining the frequency trends. We present direct evidence for these conclusions by studying the Raman modes in few layer graphene, $MoS_2$, $MoSe_2$, $WSe_2$ and $Bi_2Se_3$, using both first principles calculations and Raman spectroscopy. This study sheds light on the influence of stacking sequence on the Raman intensities of intrinsic interlayer modes in 2D layered materials in general, and leads to a practical way of identifying the stacking sequence in these materials.

**KEYWORDS**: AB and ABC stacking, few layer graphene, transition metal dicahalcogenides, $Bi_2Se_3$, Raman spectroscopy, first principles calculations, group theory

## Introduction

Two dimensional (2D) layered materials have recently attracted much attention from both research and industry communities, mainly because of their unique thickness and symmetry dependent electronic and optical properties. Among these 2D layered materials, few layer graphene (FLG), 2D transition metal dichalcogenides (TMD, such as $MoS_2$), and thin films of topological insulators $Bi_2X_3$ (X=Te, Se) are of particular interest due to their potential in novel applications. Dimensionality, symmetry and stacking orders play a critical role in the properties of layered materials. For example, 3-layer graphene (3LG) in Bernal (AB) stacking is a semimetal without a bandgap, while a spontaneous band gap can be opened in the rhombohedral (ABC) stacked 3LG with symmetry-breaking ground states[1]. ABA stacked three layer graphene exhibits unconventional quantum hall effects originating from the mirror symmetry with respect to the middle graphene layer[2]. Single layer $MoS_2$ was recently theoretically proposed and later experimentally demonstrated to be a good candidate in valleytronic applications due to the absence of inversion symmetry and large spin-orbit coupling[3, 4, 5, 6, 7]. Unlike the more common 2H type (AB stacked) $MoS_2$ with inversion symmetry, the 3R (ABC stacked) few layer $MoS_2$ is also a noncentrosymmetric material, and enhanced valley polarizations are observed on 2-4L ABC stacked $MoS_2$ in circularly polarized photoluminescence measurements[8]. More recently, the array of 2D materials available has become increasingly diverse, with new candidates such as black phosphorus, iron selenides, etc. coming into play[9, 10].

Infrared (IR) and Raman spectroscopy have become a convenient, sensitive and non-invasive procedure to characterize 2D layered materials[11, 12, 13]. By using IR absorption spectroscopy, Mak[14] and Li[15] reported unambiguous evidence for the existence of both AB and ABC stacked FLGs. On the other hand, the Raman measurements on AB and ABC stacked FLG revealed distinct line shapes for the 2D Raman mode although their frequencies are almost the same[14]. This difference in line shape is used to spatially image the different stacking orders in FLG[16, 17].

While details of Raman and IR spectra can be used to identify the stacking sequence in specific materials, the widening array of newly discovered 2D materials begs the question of whether or not there are any rules of thumb that can be used to identify the stacking sequence regardless of the specific details of the 2D material. Common to all 2D layered materials are the interlayer shear and breathing modes, in which each layer moves as a whole unit[12, 13, 18, 19, 20]. Indeed, recent experiments and calculations have uncovered such modes in AB stacked FLG [18, 22, 23, 24] and TMD[12], as well as in $Bi_2Se_3$ and $Bi_2Te_3$ (which is ABC stacked by nature)[21]. Interestingly, it was found that as the number of layers increases, the Raman frequencies of the observed interlayer shear mode blue shift in AB stacked FLG and TMD[12, 18], but red shift in ABC stacked $Bi_2Se_3$ and $Bi_2Te_3$[21]. The different frequency evolution trends arise from the different Raman intensities of the available interlayer shear modes. These trends cannot be predicted from group theory, because many Raman-active modes turn out to have zero Raman intensity[12]. Instead, we show that the frequency trends stem from Raman modes with the highest Raman intensity, which we can predict using a simple bond polarizability model that requires only information of the relative atomic positions and displacements. The model shows that stacking sequence plays a key role in determining which interlayer shear modes lead to the largest charge in polarizability (Raman intensity). Based on this model, we show that the above-mentioned correlation between stacking sequence and frequency trends is general, and can be used to determine the stacking sequence in different materials. We present direct support for these conclusions using first principles calculations as well as Raman spectroscopy measurements.

**Results**

**Interlayer modes illustrated by few layer graphene**

We begin by discussing the interlayer modes present in 2D layered materials. In a 2D material with N layers, there are N times the number of normal modes as that in the monolayer. Each normal mode in 1 layer (1L) evolves into N modes with slightly different frequencies in N layers (NL), keeping the same intralayer displacement while varying the phase difference between adjacent layers[12]. The difference in frequencies arises from the difference in relative phases in adjacent layers. In this way, the interlayer modes in NL correspond to the acoustic mode in 1L, in which all atoms within the single layer move together. Since the interlayer interactions are weak, the corresponding frequencies are also low.

As an illustration, we show in Figure 1 the ultralow frequency Raman spectra of AB and ABC stacked FLG with varying thickness, computed using density functional theory within the local density approximation (LDA) (see Methods for details). These LDA calculated frequencies are in excellent agreement with experiments (Supplementary Figure S1). The Raman intensities for each line are normalized by the largest value in that system. Unless otherwise mentioned, we shall consider in this work the Raman intensities corresponding to the most common $\bar{z}(xx)z$ polarization configuration. We use the notation "S" to label the interlayer shear modes and the subscripts, 0 to (N-1), denote in order the lowest to highest frequency shear modes in each system, with $S_0$ corresponding to the acoustic mode, $S_1$ the lowest frequency shear mode and $S_{N-1}$ the highest frequency shear mode. Supplementary Tables S1 and S2 show the frequencies of all the interlayer modes in AB and ABC stacked FLG respectively – the stacking order has negligible impact on the frequency values, indicating that the interlayer force constants are similar in both systems (see also caption of Supplementary Figure S1).   However, distinct frequency trends are found in AB and ABC stacked FLG (Figure 1b).   The peaks with largest intensity correspond to $S_{N-1}$ and $S_1$ in AB and ABC stacked FLG respectively. The other trends ($S_{N-3}$ and $S_{N-5}$ in AB stacked FLG, and $S_3$ and $S_5$ in ABC stacked FLG) are barely visible in Figure 1b, but can be seen from the details in Supplementary Tables S1 to S2.

As we have seen, the subscript of the S mode determines its relative Raman intensity.   Furthermore, this subscript, i.e. whether a mode is the lowest or highest frequency shear mode, or the third lowest or third highest frequency mode, dictates whether the frequencies red shift or blue shift with increasing thickness.   In general we find that the lower-frequency shear modes red shift with increasing thickness, and the opposite is true for the higher-frequency shear modes.   To understand this, we consider the atomic displacements of the $S_{N-1}$, $S_{N-3}$, $S_1$ and $S_3$ modes as shown in Figure 2.   First we note that the highest frequency mode $S_{N-1}$ corresponds to maximally out-of-phase displacement between adjacent layers, while the lowest frequency mode $S_1$ corresponds to minimum out-of-phase displacement.   Since the adjacent graphene layers vibrate out of phase ($180^{o}$) in the $S_{N-1}$ modes, the restoring forces accumulate with increasing thickness, resulting in the blue shift with thickness. In contrast, for the $S_1$ modes, the graphene layers are equally divided into two parts, with one part moving in one direction and the other part moving in the opposite direction, and the atomic displacements are gradually reduced towards the interface of the two parts. The relative displacement between adjacent layers therefore decreases as the number of layers increases, resulting in a red shift. Similar analysis can apply to the $S_{N-3}$ and $S_3$ modes.

For completeness, we note that besides interlayer shear modes, interlayer breathing modes are also present; the atomic displacements being similar to those of the shear modes, but in the out-of-plane direction. In FLG, the intensities of such modes are much smaller than the shear modes. We note that unlike the shear modes, the stacking order has no influence on the relative Raman intensities of the breathing modes (Supplementary Figure S1, Tables S1-S2).

The Raman activity of a phonon mode is generally assigned using group theory analysis, which we present for AB and ABC stacked FLG in the Supplementary Information. It is interesting to note that the lowest frequency shear mode, $S_1$, is Raman active in both AB and ABC stacked FLG, yet the computed Raman intensities are infinitesimally small for AB stacked FLG, but large for ABC stacked FLG (We note the detail that according to the Raman tensors (Supplementary information), there should be zero Raman intensity in the $\bar{z}(xx)z$ configuration for AB stacked N layer FLG with N odd, but possibly non-zero Raman intensity for N even). Thus, group theory cannot account for the observed frequency trends. It is interesting to note that the stacking order dependent trends for FLG are consistent with the trends for AB stacked TMD and ABC stacked $Bi_2Se_3$ and $Bi_2Te_3$ that are reported in the literature[12, 18, 21, 22, 23, 24]. We have shown, using FLG as an example, how the frequency evolution trends are determined by which shear modes (lowest or highest frequency) have the largest Raman intensities. The observation of the same trends in other materials with similar stacking order begs the question of whether we can in general conclude that the $S_1$ mode has the largest Raman intensity in ABC stacked materials, while the $S_{N-1}$ mode has the largest Raman intensity in AB stacked materials. In particular, it is interesting to ask if the same trends will be observed in AB and ABC stacked MoS$_2$ - the blue shift of the shear modes has been observed experimentally in AB stacked few layer MoS$_2$, but the space and point groups of ABC stacked FLG ( $D_{3d}^3$ ) and few layer MoS$_2$ ( $C_{3v}^1$ ) are quite different - will they have the same frequency trends? In what follows, we present a bond polarizability model that uses information about the relative atomic coordinates and atomic displacements of the vibration modes, without explicitly considering group symmetries.

**Bond Polarizability Model**

In the first principles calculations, the nonresonant Raman intensity of a phonon mode $k$ is computed in the Placzek approximation[25]:

$$I^k \propto \left| \eta \cdot \widetilde{R}^k \cdot \eta' \right|^2 \frac{(n_k + 1)}{\omega_k} \propto \frac{(n_k + 1)}{\omega_k} \left| \sum_{\alpha\beta} \eta_\alpha \eta'_\beta P_{\alpha\beta,k} \right|^2 \tag{1}$$

where $\eta$ and $\eta'$ are the unit vectors for the polarization of the incident and scattered light, $\widetilde{R}^k$ is the second rank Raman tensor, $\omega_k$ and $n_k = [\exp(\hbar\omega_k / k_B T) - 1]^{-1}$ are the frequency and the Boltzmann distribution function of phonon mode $k$, respectively. $P_{\alpha\beta,k} = \sum_{l\gamma} \left[ \frac{\partial P_{\alpha\beta}}{\partial u_{l\gamma}} \right]_0 \chi_{l\gamma}^k$ is the derivative of the electronic polarizability tensor $P_{\alpha\beta}$ with respect to the atomic displacement. $u_{l\gamma}$ is the displacement of $l$th atom in the $\gamma$ direction for normal mode $k$ and $\chi_{l\gamma}^{k}$ is the $\gamma$ th ($x$, $y$, or $z$) componenot of the eigenvector of phonon mode $k$. Here we consider the backscattering geometry with the polarization of $\eta$ and $\eta'$ parallel to the in-plane $a$ axis, i.e. $\bar{z}(xx)z$ in Porto notations[26].   From equation (1), one can deduce that for the same frequency and occupation numbers, the Raman intensity is proportional to the change in polarizability of the system, when the atoms are displaced from equilibrium in the direction of the phonon eigenvector.   The polarizability is defined as the induced dipole moment relative to the applied electic field. For the $\bar{z}(xx)z$ configuration, the relavant polarizability component would be the $xx$ component, i.e. the dipole moment in the $x$ direction, induced by an applied electric field in the $x$ direction.

As the atoms are being displaced and bonds stretched/compressed, it is reasonable to consider that the major change in polarizability will arise from changes in the relevant dipole moments of the bonds.   This has been quantified in an empirical bond polarizability model[27] which can quantitatively predict the Raman intensities of fullerene[27] and graphene ribbons[28].   In this approach, the polarizability is written as a sum of individual bond polarizabilities, which are assumed to be roughly independent of the chemical environment[28]:

$$P_{\alpha\beta} = \frac{1}{2}\sum_{l,B}\left\{\frac{1}{3}\left(\alpha_{\parallel} + 2\alpha_{\perp}\right)\delta_{\alpha\beta} + \left(\alpha_{\parallel} - \alpha_{\perp}\right)\left(R_{\alpha}R_{\beta} - \frac{1}{3}\delta_{\alpha\beta}\right)\right\}$$

(2)

where $\mathbf{R}(l,B)$ is the bond vector connecting atom $l$ to one of its nearest neighbor atoms $l'$ connected by bond $B$, the vector being normalized to unity. $\alpha_{\parallel}$ and $\alpha_{\perp}$ are the static longitudinal and perpendicular bond polarizability, respectively, which are further assumed to only depend on the bond length $R$. We note that equation (2) has further assumed cylindrical symmetry around the principal axis of each bond – although this may not be exactly correct in general systems, in this work, we are only concerned with relative magnitudes of the Raman intensities, for which these details become unimportant (see later discussion). The derivative of the bond polarizability $P_{\alpha\beta,k}$, which determines the Raman intensity, can be written as:

$$P_{\alpha\beta,k} = -\sum_{lB}\left\{\mathbf{R}_0 \cdot \vec{\chi}_l^k\left[\frac{\alpha_{\parallel}^{'} + 2\alpha_{\perp}^{'}}{3}\delta_{\alpha\beta} + \left(\alpha_{\parallel}^{'} - \alpha_{\perp}^{'}\right)\left(R_{0\alpha}R_{0\beta} - \frac{1}{3}\delta_{\alpha\beta}\right)\right] + \left(\frac{\alpha_{\parallel} - \alpha_{\perp}}{R_0}\right)\left(R_{0\alpha}\chi_{l\beta}^k + R_{0\beta}\chi_{l\alpha}^k - 2R_{0\alpha}R_{0\beta}\left(\mathbf{R}_0 \cdot \vec{\chi}_l^k\right)\right)\right\}$$

(3)

where $\mathbf{R}_0(l,B)$ is the bond vector at equilibrium configuration, normalized to unity, $R_{0\beta}$ is the $\beta$ component of $\mathbf{R}_0(l,B)$, while $R_0$ is the bond length at equilibrium. $\alpha_{\perp}^{'}$ and $\alpha_{\parallel}^{'}$ are the radial derivatives of the bond polarizability with respect to the bond length. For the $\bar{z}(xx)z$ configuration, we have:

$$P_{xx,k} = -\sum_{lB}\left\{\underbrace{\frac{\alpha_{\parallel}^{'} + 2\alpha_{\perp}^{'}}{3}\mathbf{R}_0 \cdot \vec{\chi}_l^k}_{\mathbf{I}} + \underbrace{\left(\alpha_{\parallel}^{'} - \alpha_{\perp}^{'}\right)R_{0x}^2\mathbf{R}_0 \cdot \vec{\chi}_l^k}_{\mathbf{II}} - \underbrace{\frac{1}{3}\left(\alpha_{\parallel}^{'} - \alpha_{\perp}^{'}\right)\mathbf{R}_0 \cdot \vec{\chi}_l^k}_{\mathbf{III}} + \underbrace{2R_{0x}\chi_{lx}^k\frac{\alpha_{\parallel} - \alpha_{\perp}}{R_0}}_{\mathbf{IV}} - \underbrace{2\frac{\alpha_{\parallel} - \alpha_{\perp}}{R_0}R_{0x}^2\mathbf{R}_0 \cdot \vec{\chi}_l^k}_{\mathbf{V}}\right\}$$

(4)

For the sake of clarity in explanation, we note that the RHS of equation (4) has five terms, which we will analyse in detail later.

Here, we apply the above bond polarizability model to the interlayer vibrations in layered materials. The justification for applying the bond polarizability model to the interlayer modes is shown in Figure 3, in which we find distinct charge accumulation in the interlayer regions along the axes connecting nearest neighbouring atoms, for bilayer graphene, $MoS_2$ and $Bi_2Se_3$. This indicates that although the interlayer interactions are widely believed to be of the van der Waals type, they also have some covalent character. The charge accumulated in these weak covalent bonds can produce dipole moments in the presence of an applied electric field imposed by incident radiation, resulting in Raman intensities which depend on the bond direction, i.e. the stacking order. It is because of this small amount of covalency that experimentally, different interlayer modes are observed for differently stacked materials – if there is no covalency at all, but rather completely delocalized interlayer charge distributions, the stacking order would not have such a significant impact on the Raman intensities.

Before applying the bond polarizability model in full detail, we note that we can understand quite simply the effect of stacking order on Raman intensities of the interlayer modes by using the concept of interlayer bond polarizabilities. We shall focus on the shear mode which is of interest here. For systems with in-plane isotropy, such as graphene, we can without loss of generality consider the shear modes with layers moving in the $x$ direction. When two adjacent layers move against each other, we consider how the $x$ component of the dipole moments along all nearest neighboring atoms would change. For AB and ABC stacked materials such as graphene, TMDs and $Bi_2Se_3$, each atom will have one bond $B^*$ with the largest $x$ component, and this will determine the effective direction of the $x$ component dipole moment induced by the field. In AB stacked materials, $B^*$ is pointing in opposite directions as we move from layer to layer. However, in ABC stacked materials, $B^*$ is pointing in the same direction moving from layer to layer. Therefore to maximize the change in dipole moment (for the largest Raman intensity), adjacent layers should be moving completely out-of-phase for AB stacked systems, while the whole system should be stretched out in the $x$ direction (like a deck of cards) for ABC stacked systems. This suggests that $S_{N-1}$ and $S_1$ should have the largest Raman intensities in AB and ABC stacked systems respectively, consistent with the above discussions.

We now illustrate the above qualitative picture more rigorously using equation (4) for AB and ABC stacked three layer graphene (3LG) (Figure 4). We shall sum over all interlayer bonds with bond length not larger than R for the atoms in the unit cell (the intralayer bonds will not contribute to the change in polarization since the atoms within each layer move in phase). The largest $x$ component of the interlayer bond is $a$=R*sin$\lambda$ (see Figure 4). We note that $\alpha_\perp$, $\alpha_\parallel$, $\alpha_\perp^{'}$ and $\alpha_\parallel^{'}$ are all constants. Considering the interlayer shear mode with layers moving in the $x$ direction, only the $x$ component of $\vec{\chi}_l^{k}$ is non-zero. We can now see that terms I, III and IV are proportional to $\sum_{lB} R_{0x}$ which is equal to zero from symmetry. On the other hand, terms II and V are proportional to $\sum_{lB} R_{0x}^3$ which is non-zero. For example, by summing the $x$ component of the interlayer bonds connected to atom A in the AB stacked 3LG shown in Figure 4a and 4b, we have $\sum_B R_{0x} = -\frac{a}{2R_0} - \frac{a}{2R_0} + \frac{a}{R_0} = 0$, $\sum_B R_{0x}^3 = -\frac{a^3}{8R_0^3} - \frac{a^3}{8R_0^3} + \frac{a^3}{R_0^3} = \frac{6a^3}{8R_0^3} \equiv j$. Since the B atom of the first layer is in the center of the hexagon of the second layer, both $\sum_B R_{0x}$ and $\sum_B R_{0x}^3$ will be zero, giving zero contribution for all terms in equation (4). We plot in Figure 4e-f the layer by layer top view of the interlayer bonds that will give non-zero contributions to equation (4). For AB stacked 3LG, the non-zero $\sum_B R_{0x}^3$ term is equal to $j$, -2$j$ and $j$ for atoms A, D and E respectively, while for ABC stacked 3LG, $\sum_B R_{0x}^3$ is equal to $-j$, 0 and $j$ for atoms B, C/D and E respectively.   Next to evaluate the non-zero terms II and V, we need to multiply $\sum_B R_{0x}^3 \Big|_l$ by $\chi_{lx}^k$ for each atom $l$ in the unit cell and sum over $l$.   As shown in Figure 5a, the atomic displacements of the interlayer shear modes are the same for AB and ABC stacked 3LG.   Setting $\chi_{lx}^k$ to be $\Delta_1$, $\Delta_2$ and $\Delta_3$ for atoms in the 1$^{st}$, 2$^{nd}$ and 3$^{rd}$ layers respectively, we obtain:

$$\sum_{lB} R_{0x}^3 \chi_{lx}^k = j(\Delta_1 - 2\Delta_2 + \Delta_3) \tag{5}$$

for ABstacked 3LG, and

$$\sum_{lB} R_{0x}^3 \chi_{lx}^k = j(-\Delta_1 + \Delta_3) \tag{6}$$

for ABC stacked 3LG.

For the $S_1$ mode (lowest frequency) in 3LG, $\Delta_1$=-$\Delta_3$, and $\Delta_2$=0.   From equation (5), $\sum_{lB} R_{0x}^3 \chi_{lx}^k = 0$ and therefore the Raman intensity is zero in AB stacked 3LG, while from equation (6), $\sum_{lB} R_{0x}^3 \chi_{lx}^k = -2j(\Delta_1)$ for ABC stacked 3LG.   Thus, from this model, we can explain why, although the lowest frequency modes are Raman active in both AB and ABC stacked FLG, only the lowest frequency shear mode in ABC stacked FLG has non-zero Raman intensity.   A similar analysis applies to the highest frequency $S_{N-1}$ mode.

The Raman intensities estimated from the bond polarizability model compare well with LDA calculated intensities for the interlayer shear modes of 2-7 L graphene and bulk graphite (Supplementary Figure S2).   In AB stacked bulk graphite, there is a Raman peak located around 44 cm$^{-1}$ corresponding to the $E_{2g}$ mode, but no Raman peaks are found in ABC stacked bulk graphite in the low frequency range.   In the limit of ABC stacked bulk graphite, we find that the individual bond polarizabilities for the interlayer modes will always cancel out when the periodic boundary condition is considered, thus giving zero Raman intensity.

For the breathing modes, term IV in equation (4) is zero, but in general the other terms are non-zero. However, since differences in stacking order are reflected in differences in the in-plane components of $\mathbf{R}_0(l, B)$, while it is the $z$ component of $\mathbf{R}_0(l, B)$ that shows up in equation (4), the stacking order has no influence on the relative Raman intensities of the breathing modes, consistent with our DFT calculations in the previous section.

**Generalization to other 2D materials**

Besides graphene multilayer systems, similar stacking dependent interlayer vibration modes are expected to exist in other 2D layered materials, such as $MoSe_2$, $MoS_2$, $WSe_2$ and $Bi_2Se_3$[12]. The building block of TMD is made up of three atomic layers (trilayer), with the transition metal atom covalently bonded to the chalcogen atom within the trilayer, while the building block of $Bi_2Se_3$ consists of "quintuple layers". In the interlayer vibration modes, the atoms within the building block are moving in phase with similar displacements, so the interlayer modes represent the relative displacements between the building blocks separated by the interlayer gap. Similar to graphite, the most stable structure for $MoSe_2$, $MoS_2$ and $WSe_2$ is the 2H (AB) stacked sequence, with the rhombohedral (ABC) stacked system being a metastable structure. However, for $Bi_2Se_3$, the stable structure has the rhombohedral (ABC) stacked order. The symmetry of AB stacked TMD is the same as that for AB stacked FLG, while both ABC stacked multilayer $Bi_2Se_3$ and ABC stacked FLG belong to the symmorphic space group $D_{3d}^3$ ($P\bar{3}m1$) with inversion symmetry. In contrast, ABC stacked TMD belongs to the point group $C_{3v}$ without inversion symmetry, with the irreducible representations of zone center phonon modes given by $2N(E + A_1)$.

We note that in both TMDs and $Bi_2Se_3$, the interlayer bonds are the same as those represented by the A atom in Figure 4 (except for the vertical bond between A and the atom below it). For example, each chalcogen atom has three nearest neighbouring chalcogen atoms in the adjacent layer, and from the top view, it sits in the middle of the triangle formed by these three neighbours (Figure 6 a-b). Therefore the above bond polarizability analysis for FLG also applies to these systems. Indeed, according to our DFT LDA calculations for AB and ABC stacked TMDs as well as for $Bi_2Se_3$ (ABC stacking), the shear modes with largest Raman intensity are $S_{N-1}$ and $S_1$ for AB and ABC stacked systems respectively (Figure 7). Detailed LDA results are shown in Supplementary Tables S3 to S10.

**Experimental evidence**

Next, we provide experimental evidence of the predicted trends. Besides the experimental results already published for AB stacked $MoS_2$ and $WSe_2$, and for $Bi_2Se_3$ (shown in Figure 7), we have measured the ultra-low Raman spectra for AB and ABC stacked five layer graphene (5LG) (Figure 8) as well as for AB and ABC stacked 3L $MoSe_2$ (Figure 9).

Figure 8 shows the experimental results of a mechanically exfoliated 5LG on holed 300 nm $SiO_2/Si$ substrates. The optical contrast image in Figure 8a shows that the sample is of the same thickness as 5 layer graphene. We characterized the stacking order in AB and ABC stacked FLG by comparing the Raman images of the G mode intensity and the 2D mode width[16, 17]; it is clear that there are two different domains with different stacking order in the same sample. Since the spectral band width is much larger in the ABC stacked compared with the AB stacked order in our experimental setting, we assign the much brighter part to ABC stacking, while the darker to AB stacking. The low frequency Raman spectra are then measured for the suspended sample above the hole. For the AB stacked region, the Stokes and anti-Stokes Raman spectra in Figure 8d show a typical spectral feature peak of 5LG, i.e., a shear mode is observed at 41.5 cm$^{-1}$. The Raman peak is in agreement with the reported value in other experiments[18], and in good agreement with our LDA predictions (43.2 cm$^{-1}$, corresponding to $S_{N-1}$). In contrast, our Raman spectroscopy did not detect any noticeable peak near 41 cm$^{-1}$ in the ABC stacked region, in agreement with our bond polarizability analysis and the LDA calculation. The calculation predicts the $S_1$ peak should be detected near 14 cm$^{-1}$ for the ABC stacked 5LG, however, the large background noise has merged the feature peaks of the 5LG below 15 cm$^{-1}$. Thus this experimental result has partially corroborated the LDA calculations, and is consistent with the general bond polarizability predictions.

In Figure 9, we show results from Raman scattering experiments on AB and ABC stacked $MoSe_2$ samples grown from chemical vapor deposition (CVD) [29]. Figure 9a-c shows the optical contrast and atomic force microscopy (AFM) image of a 3L $MoSe_2$ sample; the measured Raman spectra on different 3L $MoSe_2$ samples are shown in Figure 9d.    Interestingly, 35% of all 43 samples exhibit only a peak around 13.3cm$^{-1}$, very close to the theoretical prediction of 15.1cm$^{-1}$ for the lowest frequency shear mode $S_1$ in 3L ABC stacked MoSe$_2$ samples, while 28% of all samples exhibit a peak at 23.1cm$^{-1}$, close to our theoretically predicted value of 22.1cm$^{-1}$ for the highest frequency shear mode $S_{N-1}$ in 3L AB stacked MoSe$_2$.    Since exfoliated 2H MoSe$_2$ has the AB stacking order, we compare our results with Raman spectra on the exfoliated sample.    The Raman peak of the exfoliated AB stacked sample is around 23.3 cm$^{-1}$, confirming that the samples that show a peak around 23.1cm$^{-1}$ are AB stacked.    It is noted that 37% of all samples detect both $S_1$ and $S_{N-1}$ modes, because the spot size of our laser is at least 1μm, and the 2H and 3R phase can coexist in the same sample with a sharp few-nanometer wide transition boundary[29].    Furthermore, scanning tunneling electron microscopy (STEM) experiments performed on the CVD-grown samples[29] confirmed that ABC stacking was more prevalent, consistent with our assignment of stacking orders above.    Experimental data for MoSe$_2$ thin films with different number of layers is plotted in Figure 7c; the results agree well with the LDA results.

**Discussion**

The advantage of a bond polarizability model is that it can easily be applied to more complicated stacking sequences, such as the ABACA stacking in 5 layer MoSe$_2$. We show these predictions in Figure 5c.    As the ABACA stacking order could not be distinguished in STEM, the predicted Raman intensities along with experimental Raman spectra were used to assign the stacking sequence in a few of our CVD-grown 5 layer MoSe$_2$ samples[30]. We also note that the bond polarizability predictions are consistent with our LDA results for these complicated stacking sequences. The above results provide evidence that although the space groups are not all the same for materials with the same stacking order, the frequency evolution trends as predicted by our bond polarizability model are general. Furthermore, these predictions can be used to determine the stacking orders in experimental samples, even when the stacking order cannot be distinguished by microscopy.

Finally, we note that in few layer black phosphorus (Figure 6d), the in-plane shear modes cannot be detected under the $\bar{z}(xx)z$ polarization configuration. Considering both shear modes in the $x$ and $y$ directions, we can see that any change in bond polarizability will cancel out when we sum over all nearest neighbor bonds, because of the mirror planes along the $y$ and $x$ directions respectively. On the other hand, the breathing modes can have non-zero Raman intensity in the $\bar{z}(xx)z$ configuration. These predictions are consistent with first principles calculations and Raman spectroscopy experiments[31].

In conclusion, we have shown that the interlayer bond polarizabilities can allow us to understand, both intuitively and semi-quantitatively, the Raman intensities of interlayer modes in general 2D layered materials. Specifically we find that the change in polarizability is maximized for the lowest frequency shear mode in ABC stacked materials, but for the highest frequency shear mode in AB stacked materials, regardless of the details of the space group. The resultant differences in Raman intensity result in clear and distinct frequency trends for AB and ABC stacked systems

– as the number of layers increases, the interlayer shear mode red shifts for ABC stacked systems, and blue shifts for AB stacked systems. Because these trends are distinct, and furthermore, do not overlap, they provide a general way to distinguish AB and ABC stacking in 2D layered materials. This bond polarizability model also provides consistent results when applied to interlayer modes in other materials such as black phosphorus, and can be used as a tool to make quick predictions on Raman intensities for more complicated stacking orders. Our predictions are substantiated with extensive first principles calculations as well as Raman spectroscopy measurements on different 2D materials.

## Methods

First principles calculations of vibrational Raman spectra are performed within density-functional perturbation theory (DFPT) as implemented in the plane-wave code QUANTUM-ESPRESSO[32]. The local density approximation (LDA)[33] to the exchange-correlation functional with projector-augmented wave potentials is employed for the calculation of phonon frequencies, while the non-resonant Raman intensity is calculated using the norm-conserving pseudopotential, within the Placzek approximation[25]. With the frequency and Raman intensity, a scale parameter of 1 is used in the Lorentzian broadening function to get the calculated Raman spectrum. To get the converged results, a plane-wave kinetic energy cutoff of 65 Ry is used for the wave functions, and the convergence threshold is set to $10^{-9}$ eV and $10^{-18}$ eV in the electron and phonon self-consistent calculation, respectively. The structures are considered as relaxed when the maximum component of the Hellmann-Feynman force acting on each atom is less than 0.003 eV/Å. A Monkhorst-Pack k-point mesh of 44×44×1, 17×17×1 and 11×11×1 are used to sample the Brillouin Zones for the FLG, TMD and $Bi_2Se_3$ systems, respectively. We use a vacuum thickness of 16 Å in the direction perpendicular to the slabs to prevent interactions between periodic slab images. The spin-orbit coupling effect is included self-consistently by using fully relativistic pseudopotentials for the valence electrons in $Bi_2Se_3$.

Experimentally, the graphene layers and $MoSe_2$ are prepared by the mechanical exfoliation method[34] and chemical vapor deposition method[29], respectively. The Raman spectroscopy measurements are conducted in a backscattering geometry, excited with a Helium-Neon laser with λ = 532 nm for $MoSe_2$ and graphene layers. The detection of ultralow frequency is achieved by filtering out the laser side bands through the adoption of a reflecting Bragg grating, the ultralow frequency of $MoSe_2$ is achieved by triple-grating setup (Horiba-JY T64000). The laser power is kept below 0.05mW on the sample surface to avoid laser-induced heating.

## Acknowledgments

S.Y.Q and X.L gratefully acknowledges A*STAR for funding under IHPC Independent Investigatorship, and the Singapore National Research Foundation (NRF) for funding under the NRF Fellowship (NRF-NRFF2013-07). T.Y. acknowledges Singapore NRF under NRF RF Award No. NRF-RF2010-07 and MOE Tier 2 MOE2012-T2-2-049. Q.X. thanks Singapore NRF *via* a Fellowship grant (NRF-RF2009-06) and an Investigatorship grant (NRF-NRFI2015-03), MOE *via* a tier2 grant (MOE2012-T2-2-086). The computations were performed on NUS Graphene Research Centre and the A*STAR Computational Resource Center. S.Y.Q. thanks K.H. Khoo for first bringing her attention to the bond polarizability model.

## Author Contributions

X.Luo performed all the calculations including earlier calculations which inspired this work. S.Y.Q. conceived the idea. X.Luo and S.Y.Q. performed the theoretical analysis and wrote the main manuscript text. C.C. and T.Y. contributed the experimental data on graphene. X.Lu and Q.X. contributed the experimental data on $MoSe_2$ as well as earlier experiments that inspired this work. All authors reviewed the manuscript.

## Additional information

**Competing financial interests:** The authors declare no competing financial interests.

**Supplementary information** accompanies this paper at http://www.nature.com/scientificreports

**Figure Legends**

**Figure 1**. (a) Structural illustrations of AB and ABC stacked order in 3L graphene (extended periodically in the a and b directions). Pink and gray balls denote atoms in the two sublattices. (b) LDA calculated low-frequency Raman spectra in AB and ABC stacked few layer graphene (FLG) and bulk graphite. Dashed lines guide the frequency evolutions of the interlayer shear modes.

**Figure 2**. The atomic displacements and frequencies of the $S_{N-1}$ and $S_{N-3}$ modes in AB stacked FLG and $S_1$ and $S_3$ modes in ABC stacked FLG.

**Figure 3**. LDA and vdW calculated charge density difference for bilayer (a) graphene, (b) $MoS_2$ and (c) $Bi_2Se_3$ at the equilibrium distance. The charge accumulation and depletion is denoted by the yellow and blue color, respectively.

**Figure 4**. Top view and side view of the (a) (b) AB stacked 3LG and (c) (d) ABC stacked 3LG. The gray and brown balls respectively represent the C atoms in the odd and even numbered layers in AB stacked 3LG, while gray, brown and blue balls represent C atoms in the first, second and third layer of ABC stacked 3LG, respectively. (e) and (f) show the top view of the projected bonds for each layers. The solid and dash lines refer respectively to bonds connected downward and upward. The unit cell of 3LG is shown by the blue diamond.

**Figure 5**. (a) Atomic displacements of the interlayer shear modes in AB and ABC stacked 3LG. (b) General bond polarizability model in AB and ABC stacked 3LG, a sphere represents a graphene layer. (c) The application of general bond polarizability model to ABACA stacked 5 layer $MoSe_2$.

**Figure 6**. Top view and side view of (a) AB stacked (b) ABC stacked $MoS_2$, (c) $Bi_2Se_3$ and (d) Black phosphorus. To have a better visual effect, the S atoms in odd and even numbered layers are shown in grey and brown in AB stacked $MoS_2$, while in ABC stacked $MoS_2$, they are shown in grey, brown, and blue for $1^{st}$, $2^{nd}$ and $3^{rd}$ layers respectively. The P atoms near the interlayer gap are plotted in grey and blue. The interlayer gaps are denoted by black double arrows.

**Figure 7**. Frequency evolutions of interlayer shear modes with largest Raman intensity in (a) $MoS_2$, (b) $WSe_2$, (c) $MoSe_2$ and (d) $Bi_2Se_3$ with different stacking orders. The LDA calculated data are plotted with solid dots and the experimental data are shown in empty dots. The gray and indigo lines result from fitting of the largest frequency shear modes ($S_{N-1}$) and the lowest frequency shear modes ($S_1$) for AB and ABC stacking order, respectively, using the linear chain model.[12] The resulting force constants are shown in Table S10.

**Figure 8**. (a) Optical images of a 5 layer graphene (5LG) sample and the corresponding Raman images of (b) the G mode intensity and (c) the 2D mode linewidth. (d) Experimentally measured Raman spectra of 5LG.

**Figure 9**. (a) Optical and (b) AFM image of a 3L $MoSe_2$ sample, the cross cection height is shown in (c). (d) Raman spectra of the 3L $MoSe_2$ with different stacking orders.

**Figure 1**

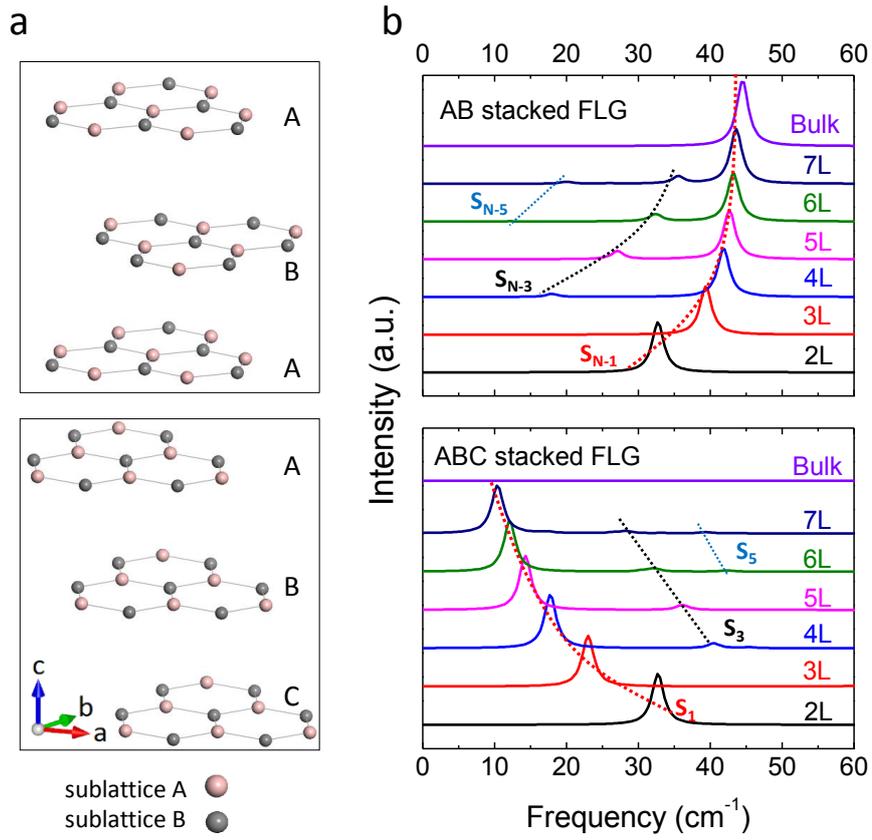

**Figure 2**

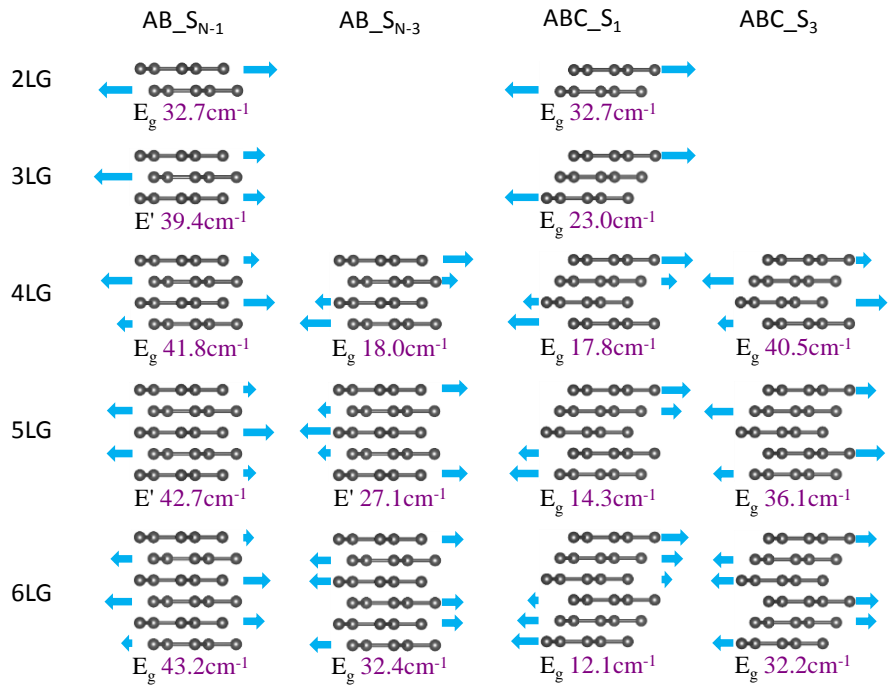

**Figure 3**

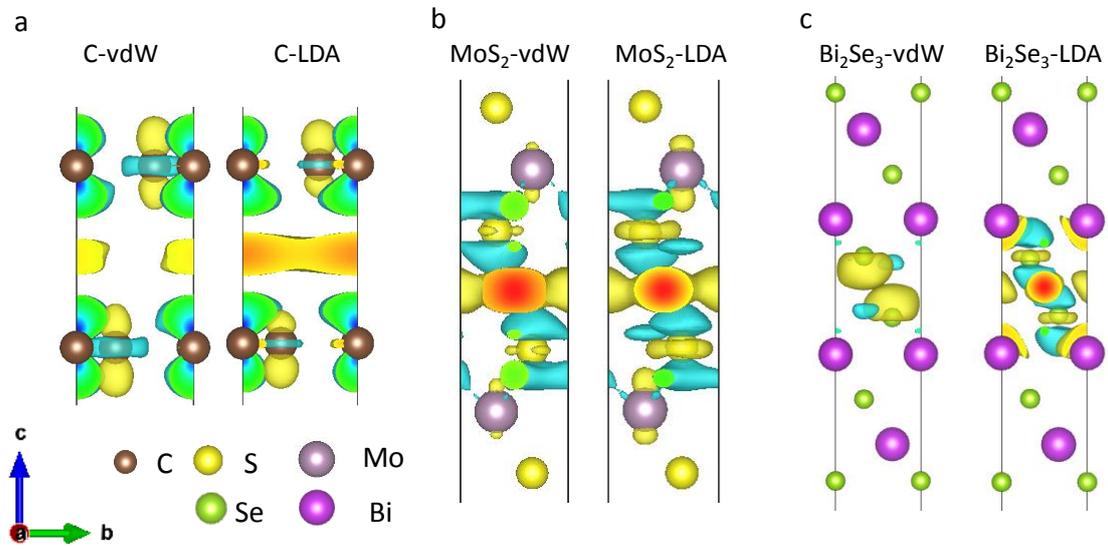

a   C-vdW   C-LDA

b   MoS₂-vdW   MoS₂-LDA

c   Bi₂Se₃-vdW   Bi₂Se₃-LDA

c

b

a

C
Se
S
Bi
Mo

**Figure 4**

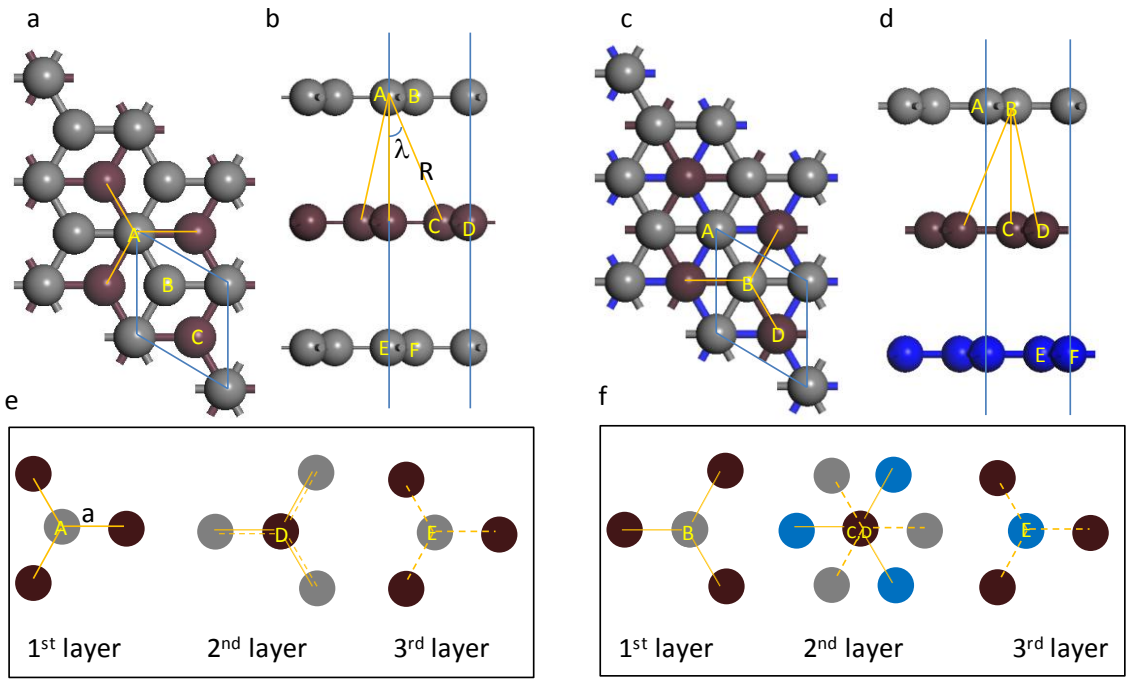

a

b

c

d

e   1st layer   2nd layer   3rd layer

f   1st layer   2nd layer   3rd layer

**Figure 5**

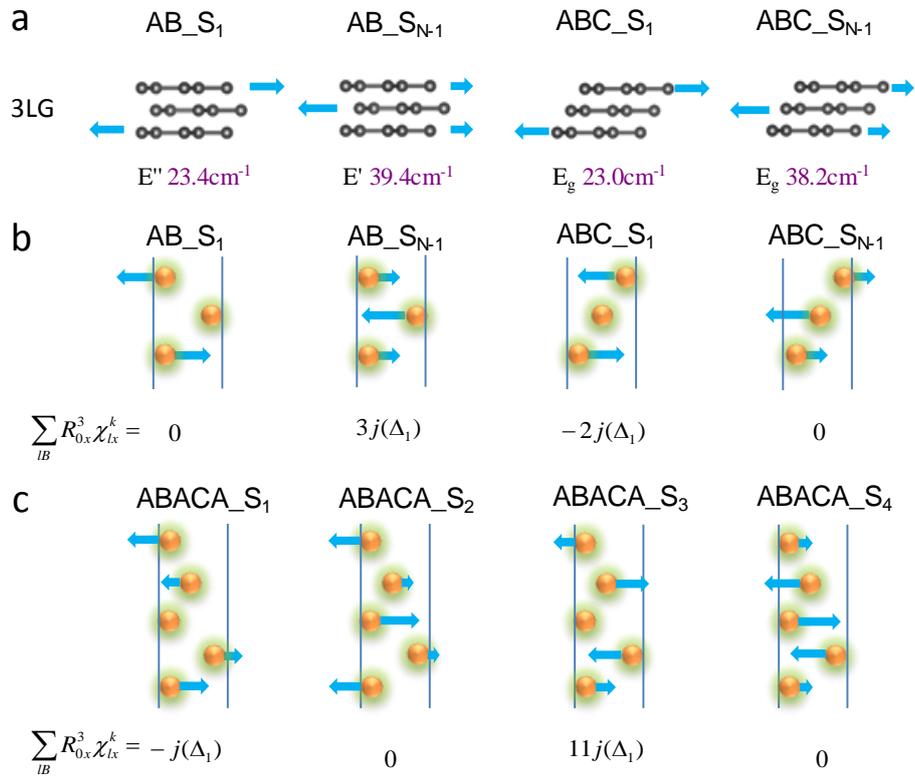

a

3LG

AB_S$_1$      AB_S$_{N-1}$      ABC_S$_1$      ABC_S$_{N-1}$

E'' 23.4cm$^{-1}$      E' 39.4cm$^{-1}$      E$_g$ 23.0cm$^{-1}$      E$_g$ 38.2cm$^{-1}$

b

AB_S$_1$      AB_S$_{N-1}$      ABC_S$_1$      ABC_S$_{N-1}$

$\sum_{IB} R_{0x}^3 \chi_{lx}^k =$  0      $3j(\Delta_1)$      $-2j(\Delta_1)$      0

c

ABACA_S$_1$      ABACA_S$_2$      ABACA_S$_3$      ABACA_S$_4$

$\sum_{IB} R_{0x}^3 \chi_{lx}^k = -j(\Delta_1)$      0      $11j(\Delta_1)$      0

**Figure 6**

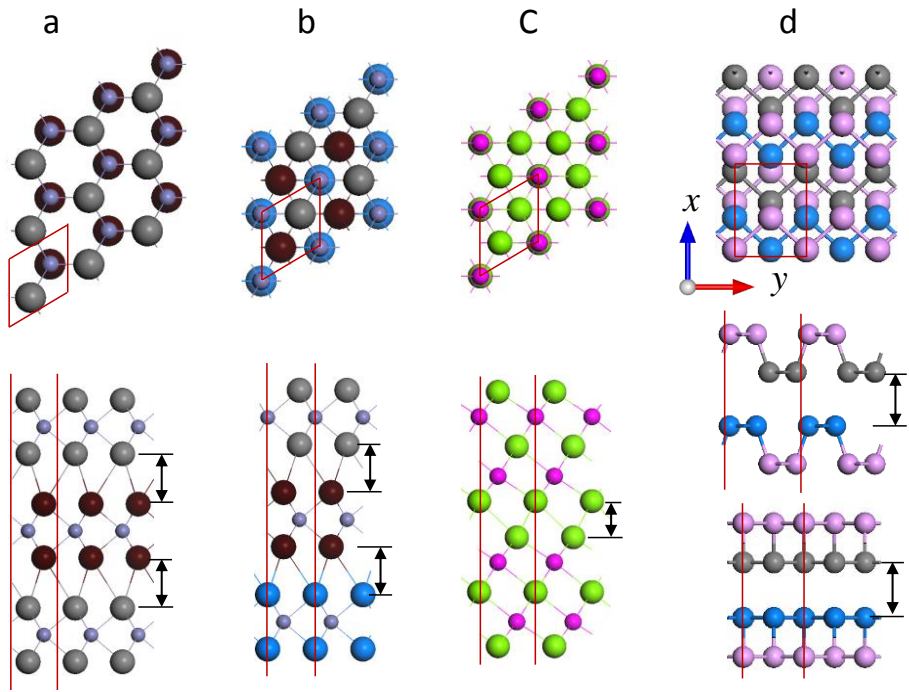

a          b          C          d

$x$
$y$

## Figure 7

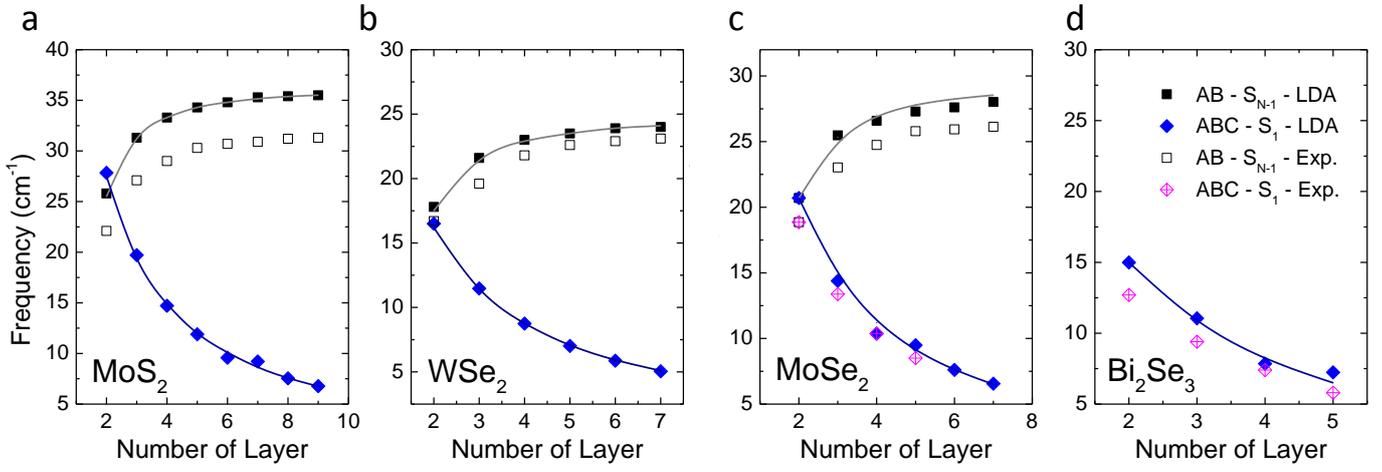

## Figure 8

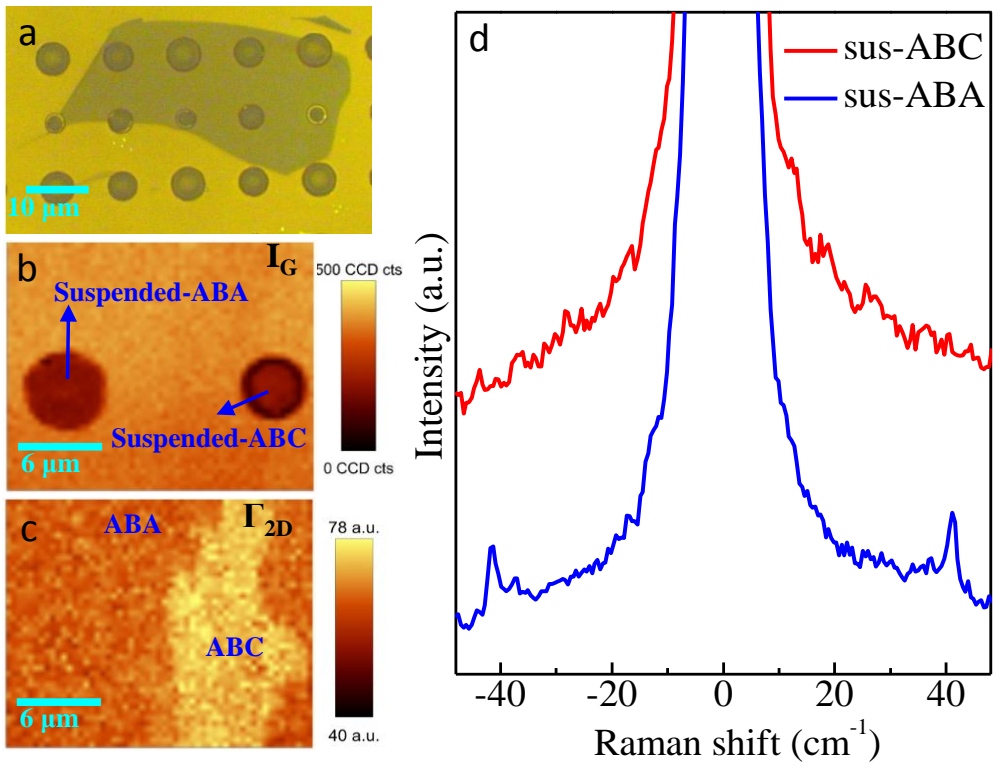

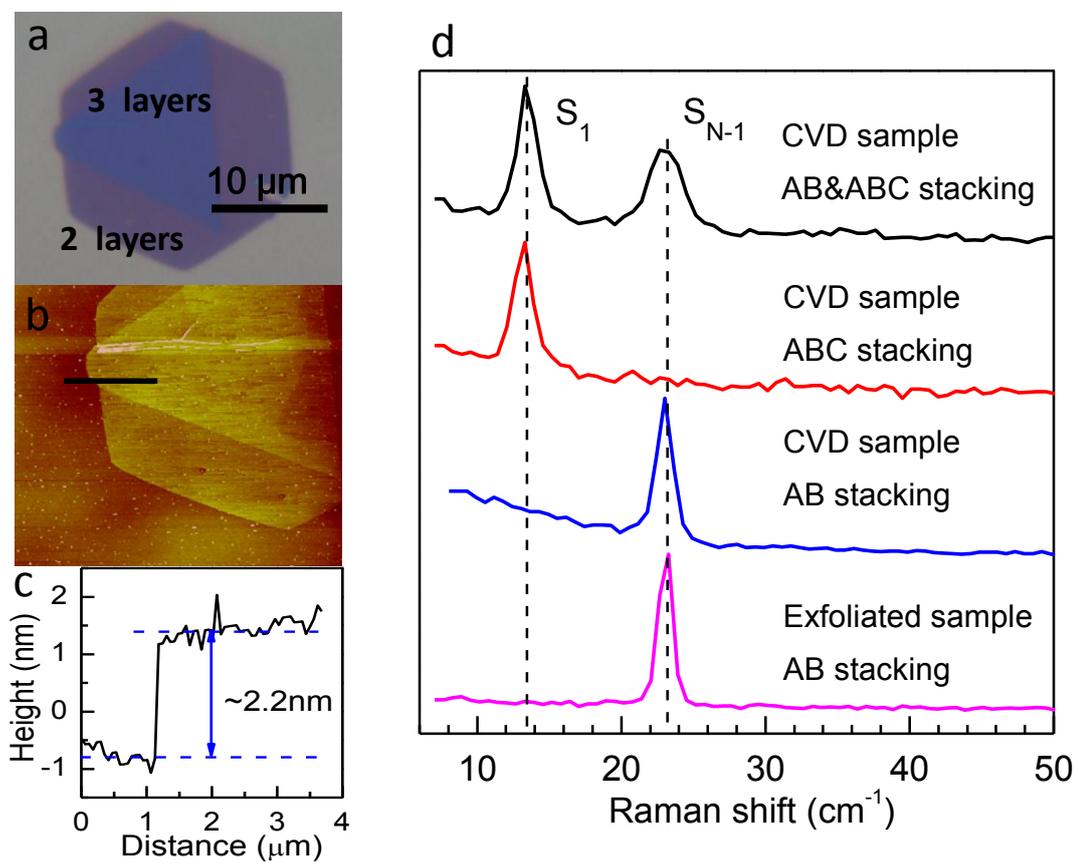